\documentclass{article}  
\usepackage{breckenr2005}
\usepackage{graphicx}
\usepackage{bm}
\usepackage{wrapfig}
\usepackage{amssymb}
\frompage{000} \topage{000}                                              
\def\rightmark{Single--spin Asymmetries in Semi--inclusive Pion Production}
\def\leftmark{U.~Elschenbroich}

\newcommand{\up}{\uparrow}
\newcommand{\down}{\downarrow}
\newcommand{\lmom}{\bm{\langle}}
\newcommand{\rmom}{\bm{\rangle}}
\renewcommand{\vec}{\bm}

\title{Single--spin Asymmetries in Semi--inclusive Pion Production} 
\authors{
{Ulrike Elschenbroich$^{1a}$ (on behalf of the \textsc{Hermes} Collaboration)%
}\\[2.812mm]
{\normalsize
\hspace*{-8pt}$^1$ Department of Subatomic and Radiation Physics, \\ 
University of Ghent, Belgium\\[0.2ex] 
}}
 
\abstract{For the first time single--spin asymmetries in semi--inclusive pion production
  are measured by the \textsc{Hermes} experiment with a transversely polarised hydrogen 
  target. Two different sine--modulations are extracted which can be related to the 
  \textit{transversity} $\delta q(x)$ and \textit{Sivers}  $f_{1T}^{\perp q}(x)$ 
  quark distribution functions. The extracted sine--moments still contain small 
  sub--leading twist contributions which can be extracted from \textsc{Hermes} data 
  combining the results from the transversely polarised hydrogen target with previously 
  measured results from a longitudinally polarised hydrogen target.}

 
\begin{document}
 
\maketitle
\setcounter{page}{1}

\section{Introduction}\label{intro}

For the last decades deep inelastic scattering (DIS) was and still is a very successful 
tool to investigate the structure of the nucleon. In this process a lepton scatters off 
a nucleon via the exchange of a single virtual photon. A measure of the spatial 
resolution in the scattering process is the four--momentum transfer to the target 
nucleon. In DIS processes this momentum transfer is large enough to resolve the 
constituents of the nucleon.

In the quark parton model in which the virtual photon is assumed to scatter
incoherently off the quarks in the nucleon, the DIS cross section can be expanded
in terms of \textit{quark distribution functions}. 
The leading--twist distribution functions (DF) can be interpreted as probability 
densities dependent on the longitudinal quark momentum in a frame in which the nucleon 
is moving towards the photon with ``infinite'' momentum.
When integrating the leading--twist DFs over the intrinsic transverse quark momentum 
$p_T$, only three of them remain. Two of these DF have already been explored for 
different quark flavours $q$ by several experiments \cite{Mar02,Air04}.
These are the unpolarised DF $q(x)$ and the helicity DF $\Delta q(x)$, which gives the 
probability to find a quark with its helicity parallel to the nucleon helicity.
Here $x$ is the dimensionless Bjorken scaling variable. It can be identified with the 
fractional momentum of the nucleon carried by the struck quark.
The last of the three DFs is the chiral--odd \textit{transversity} 
\cite{Ral79,Art90,Jaf92} function $\delta q(x)$.
In the basis of transverse spin eigenstates it has the simple interpretation of the degree 
to which the quarks are polarised along the proton's spin direction when the proton is 
polarised transversely to the virtual photon.

\begin{figure}[t]
  \hspace*{0.5cm}
  \includegraphics[height=6.2cm,angle=270]{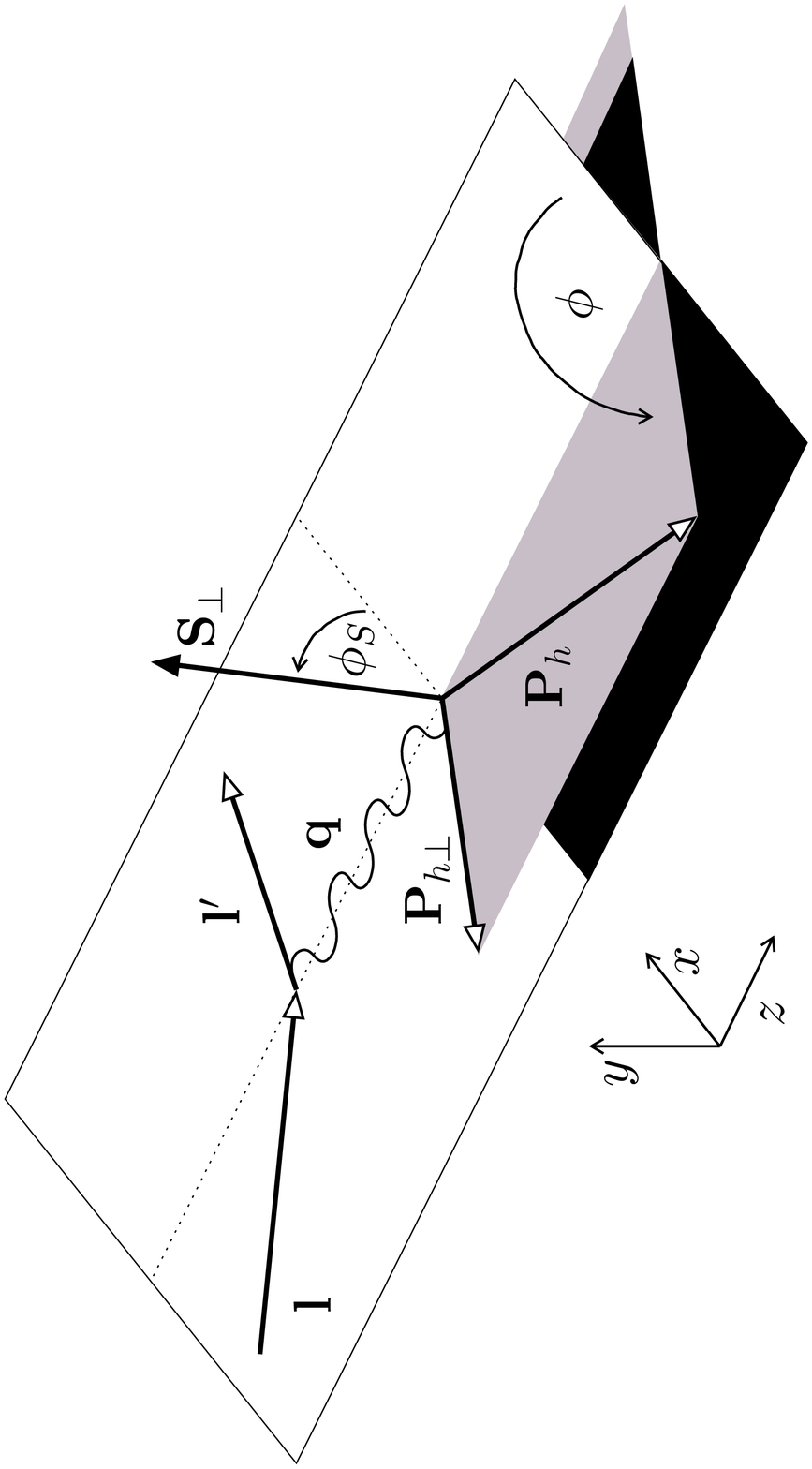}

  \vspace*{-3.4cm}\hspace*{6.2cm}
  \includegraphics[height=5.8cm,angle=90]{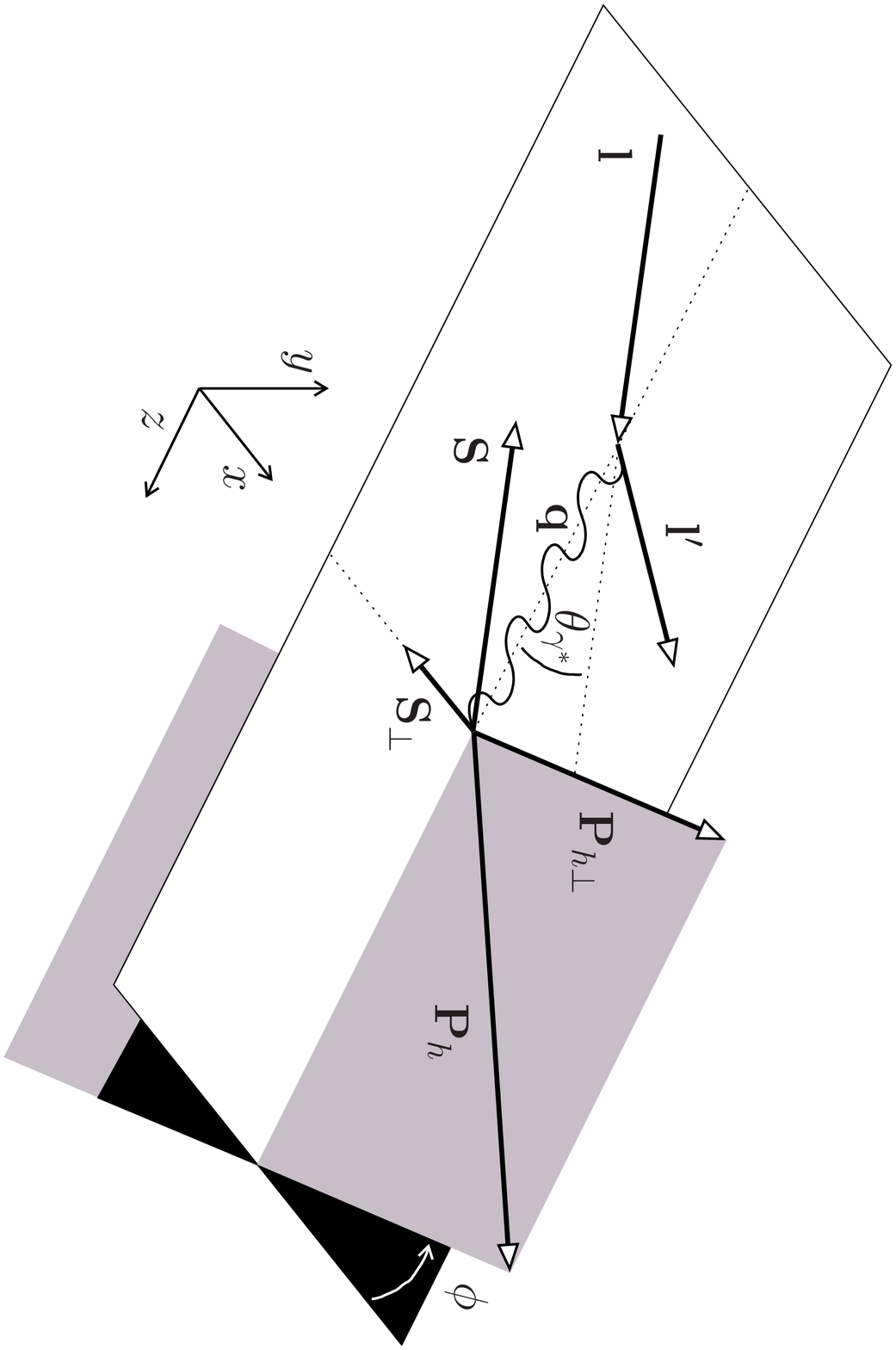}
  \caption[]{Schematic illustration of the deep inelastic scattering process off a 
    transversely (left) and longitudinally (right) polarised target.}
  \label{phiDefinition}
\end{figure}

Since DIS is a hard scattering process in which chirality is conserved, 
transversity can only be measured combined with another chiral--odd object.
This can be a fragmentation function (FF) which appears in addition to the DFs in the 
cross section of semi--inclusive DIS when produced hadrons are detected in addition to 
the scattered lepton. In unpolarised semi--inclusive DIS for instance the cross section 
is proportional to the product of the unpolarised DF $q(x)$ and the unpolarised FF
$D_1^q(z)$ which gives the probability density that a struck quark of flavour
$q$ produces a certain final state hadron with the fractional energy $z$.
In case of a transversely polarised nucleon target, the transversity distribution 
enters the cross section combined with the chiral--odd FF $H_1^{\perp q}(z)$ known as
\textit{Collins} function \cite{Col93}. In addition a second DF $f_{1T}^{\perp q}$ appears
in the cross section together with the unpolarised FF. This DF -- which is called 
\textit{Sivers} function  \cite{Siv90} -- relates the quark transverse momentum with the 
transverse polarisation of the nucleon.
The property of the Sivers function to be odd under time reversal (T--odd)
was believed to forbid its existence. But recently it was realised that
final--state interactions via a soft gluon offer a mechanism to create the
necessary interference of amplitudes \cite{Bro02} for the existence of the
so--called ``na\"ive T--odd'' nature of the Sivers function. For DFs in DIS 
``na\"ive T--odd'' means time--reversal odd neglecting final-state interaction. 
An interesting consequence of a non--zero Sivers function is the existence of 
orbital angular momentum of the quarks \cite{Bro02,Bur04}.

\section{Azimuthal Asymmetries}

Since Sivers and Collins functions do not survive integration over
the intrinsic transverse momentum $p_T$ and the transverse momentum $k_T$
acquired in the fragmentation process, respectively, the tools to measure
the objects of interest are azimuthal asymmetries. These asymmetries depend on
two azimuthal angles $\phi$ and $\phi_S$ drawn in Fig.~\ref{phiDefinition}.
The angle  $\phi_S$ is the angle between the lepton scattering plane containing the
incoming and outgoing lepton and the transverse spin component of the target nucleon.
The hadron angle $\phi$ is defined between the lepton scattering plane and the hadron 
production plane spanned by the produced hadron and the virtual photon.

When including azimuthal dependencies the cross section can be expanded in terms of 
different sine and cosine modulations of $\phi$ and $\phi_S$. It contains
among others:
\begin{eqnarray}
  \sigma(x,y,z,\phi,\phi_S) &\sim& \dots S_\perp \sin(\phi+\phi_S) \sum_q e_q^2 
  \delta q(x) H_1^{\perp(1/2)q}(z)\ + \nonumber\\
  && \dots S_\perp \sin(\phi-\phi_S) \sum_q e_q^2 f_{1T}^{\perp(1/2)q}(x) D_1^q(z)\ ,
  \label{eq:sigma}
\end{eqnarray}
where $S_\perp$ is the transverse target polarisation with respect 
to the virtual photon and $e_q$ is the charge of the different quark flavours $q$.
The superscript $(1/2)$ denotes the $|p_T|$-- or $|k_T|$--moment of the DF or FF, respectively. 
Gaussian distributions are assumed for $p_T$ and $k_T$ in order to solve the convolution 
integrals in which the products of DF and FF appear \cite{Mul96}.

Experimentally the various terms in the cross section can be extracted by measuring azimuthal
single--spin asymmetries and picking up the interesting sine modulations. For the two terms
in Eq.~(\ref{eq:sigma}) one needs a polarised target. The luminosity normalised 
count rate asymmetry between opposite target spin states ($\up$,$\down$) reads:
\begin{eqnarray}
  A_\mathrm{UT}(\phi,\phi_S)\!\!\! &=& 
  \!\!\!\frac{1}{S_\perp}\frac{N^\up(\phi,\phi_S)-N^\down(\phi,\phi_S)}%
  {N^\up(\phi,\phi_S)+N^\down(\phi,\phi_S)}\nonumber\\
  &=&\!\!\! 2 \lmom \sin(\phi+\phi_S)\rmom_\mathrm{UT} \sin(\phi+\phi_S) + 
  2 \lmom \sin(\phi-\phi_S)\rmom_\mathrm{UT} \sin(\phi-\phi_S)\ ,\qquad
\end{eqnarray}
where the subscript UT indicates an unpolarised lepton beam and a transversely polarised
target. The asymmetry moments $\lmom \sin(\phi\pm\phi_S)\rmom_\mathrm{UT}$ are
extracted using a two--dimensional fit.

\section{The HERMES Experiment}

The \textsc{Hermes} experiment at \textsc{Desy} uses the 27.5 GeV positron beam
provided by the \textsc{Hera} storage ring. The beam interacts with an internal hydrogen 
gas target \cite{Air05} which was longitudinally polarised with an average polarisation
of $0.83\pm0.04$ (systematic) in the years 1996 and 1997.
In the year 2002 the polarisation was rotated to achieve a transversely polarised target
with an average polarisation of $0.78\pm0.04$ (systematic).
The scattered positrons and the produced hadrons are detected with the \textsc{Hermes} 
spectrometer \cite{Ack98}.
This spectrometer provides lepton identification with an average efficiency of 98\% at a 
hadron contamination of less than 1\%. The gas threshold \v Cerenkov detector which was used for 
pion identification was replaced by a Ring-Imaging \v Cerenkov detector (\textsc{Rich}) in 1998 
which allows the efficient identification of charged pions, kaons, and 
protons over almost the complete momentum range and hence leads to a very clean pion 
sample.

\begin{wrapfigure}{r}{6cm}
  \includegraphics[width=6cm]{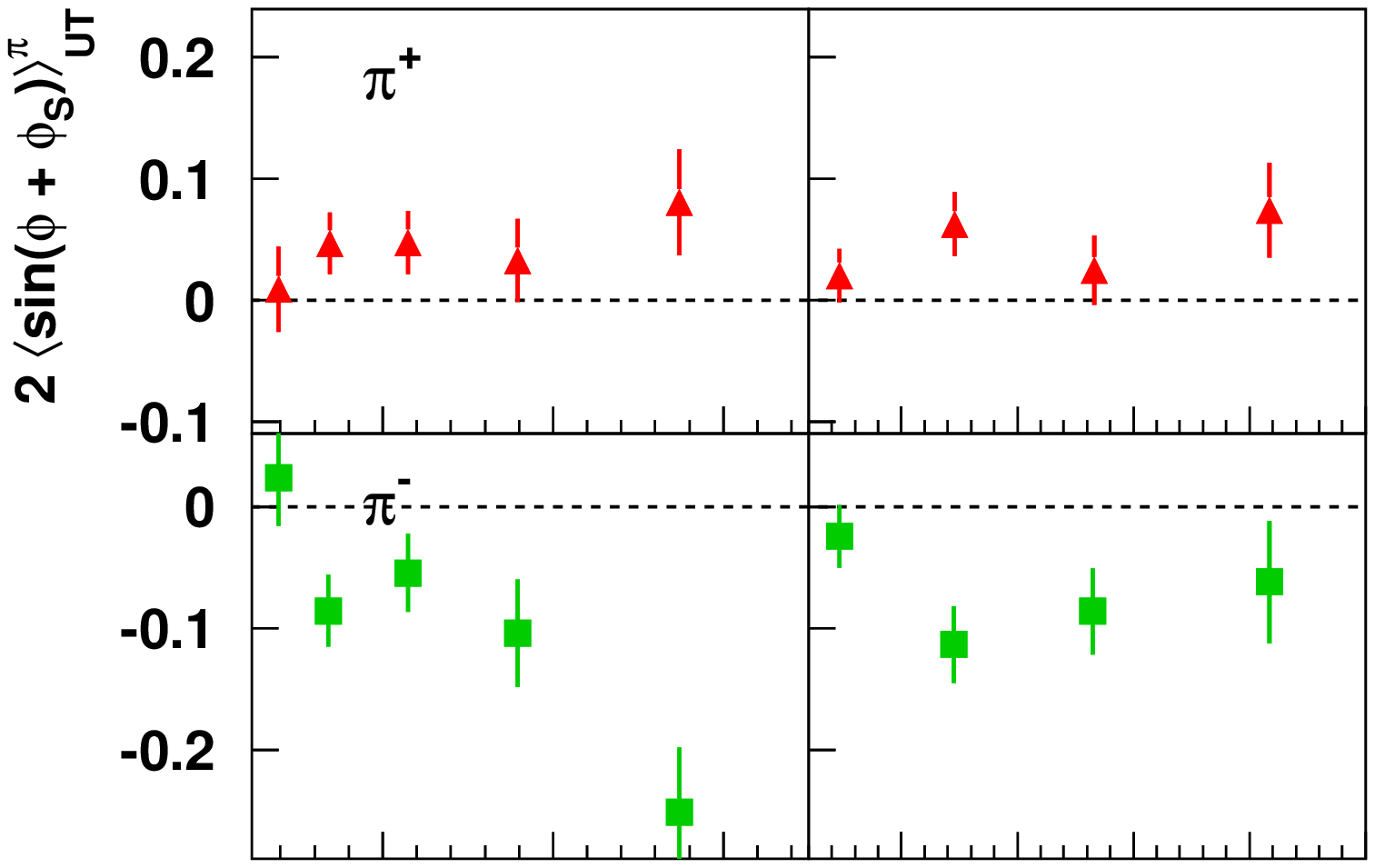}\\
  \includegraphics[width=6cm]{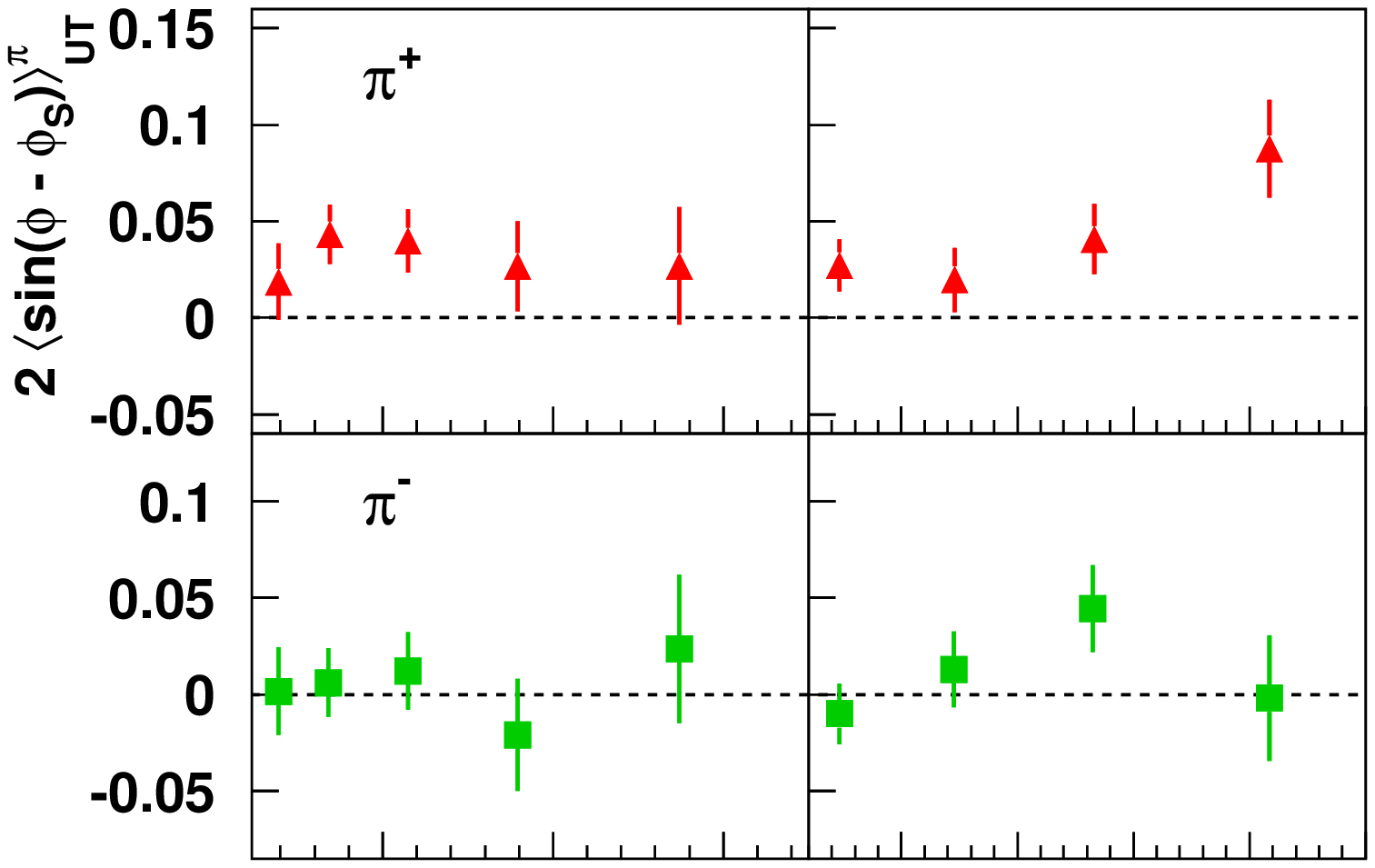}\\
  \hspace*{0.2cm}\includegraphics[width=5.8cm]{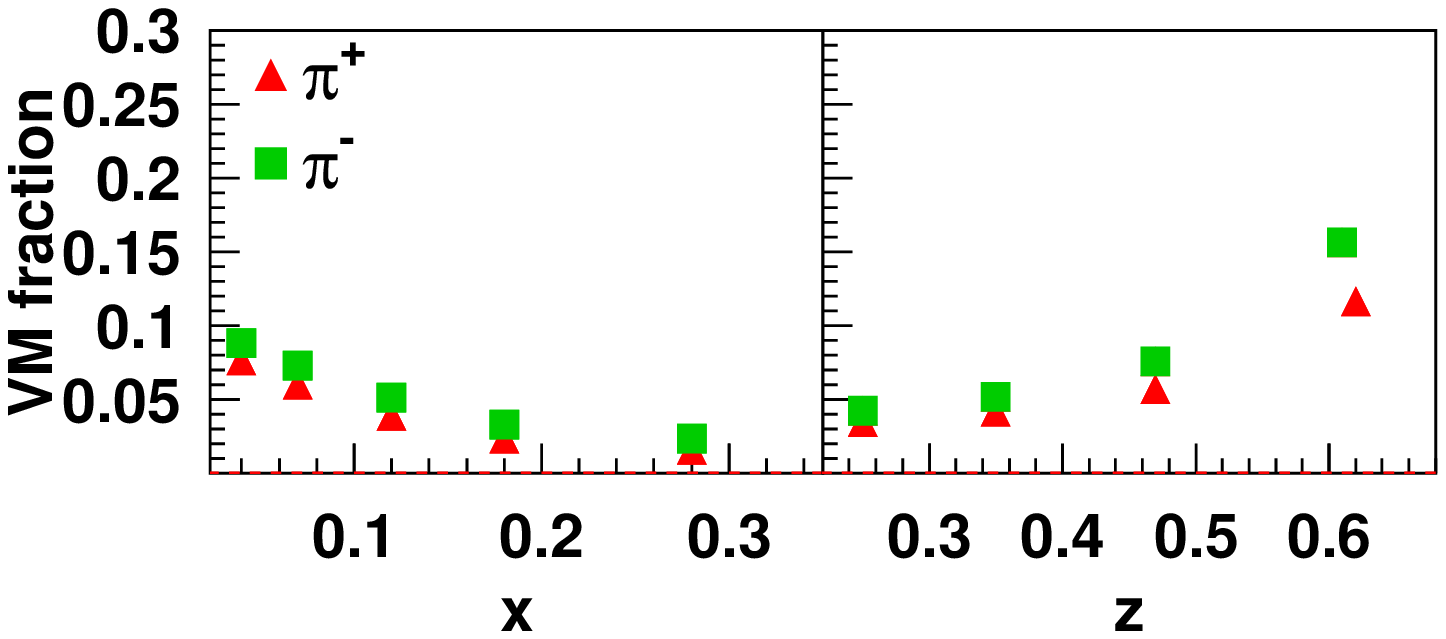}
  \\[-0.7cm]
\caption[]{Measured asymmetry moments\\
  $2\lmom\sin(\phi\pm\phi_S)\rmom_\mathrm{UT}$
  for charged pions in\\ different bins in $x$ and $z$.}
\label{AUT}
\end{wrapfigure}
\section{Transverse Target Results}

In Fig.~\ref{AUT} the measured asymmetry moments 
$2\lmom\sin(\phi\pm\phi_S)\rmom_\mathrm{UT}$ for charged pions are plotted
\cite{Air05_2}, where in the left (right) column the dependency on the kinematic variable $x$ 
($z$) is shown. In addition to the plotted statistical uncertainties there is an overall
8\% scale uncertainty in the moments dominated by the uncertainty of the target polarisation.
The asymmetry moments $\lmom\sin(\phi+\phi_S)\rmom_\mathrm{UT}$
containing transversity and the Collins FF include an additional kinematic factor 
(for details see \cite{Air05_2}). These moments are positive for positive pions and 
negative for negative pions, consistent with the expectation of a positive transversity 
$\delta u$ for $u$--quarks and a negative $\delta d$. Astonishing is the larger absolute 
value of the moments for $\pi^-$ compared to the moments for $\pi^+$. This could be 
explained by a disfavoured Collins function with the opposite sign and the same magnitude
as the favoured Collins function. 
The asymmetry moments $\lmom\sin(\phi-\phi_S)\rmom_\mathrm{UT}$ are compatible with 
zero for negative pions but significantly positive for $\pi^+$. This is the first hint of
a ``na\"ive T--odd'' DF from DIS. The data samples include pions coming from decays of 
diffractively produced vector mesons. The contribution estimated with the \textsc{Pythia6} 
event generator \cite{Sjo01} is shown in the lowest panel.

\section{Longitudinal Single--Spin Asymmetries Revisited}

The \textsc{Hermes} experiment has also published single--spin asymmetries on 
longitudinally polarised hydrogen \cite{Air00}. For this measurement the nucleons were 
polarised longitudinally with respect to the incoming lepton beam under an angle 
$\theta_{\gamma^*}$ to the direction of the virtual photon ($\vec q$) 
(see Fig.~\ref{phiDefinition}).
Therefore both polarisation components $S_\|$ and $S_\perp$ with respect to $\vec q$ exist
where the transverse component is suppressed by the factor $\sin\theta_{\gamma^*}$.
In the following, the measured asymmetry moments with
respect to the lepton beam get a primed subscript $\mathrm{UL^\prime}$ or 
$\mathrm{UT^\prime}$, respectively. Asymmetry moments with unprimed subscripts are defined 
with respect to the virtual photon.
Since the angle $\phi_S$ has the constant value $\pi$, the moments 
$\lmom \sin(\phi\pm\phi_S)\rmom_\mathrm{UT}$ reveal themselves as a $\sin\phi$
modulation of the cross section, proportional to the small polarisation component $S_\perp$. 
Also in the cross section for scattering off longitudinally
polarised nucleons with respect to the virtual photon ($S_\|$) a $\sin\phi$ modulation exists.
The corresponding moment $\lmom \sin\phi \rmom_\mathrm{UL}$ is a subleading--twist 
contribution in contrast to $\lmom \sin(\phi\pm\phi_S)\rmom_\mathrm{UT}$ which 
involves only leading--twist DFs and FFs \cite{Bac04}. Due to the suppression of 
subleading--twist contributions the \textsc{Hermes} results have been interpreted in terms 
of the Collins or Sivers functions neglecting some or all the contributions from the 
longitudinal polarisation component.

\begin{figure}[t]
  \centerline{\includegraphics[width=7cm]{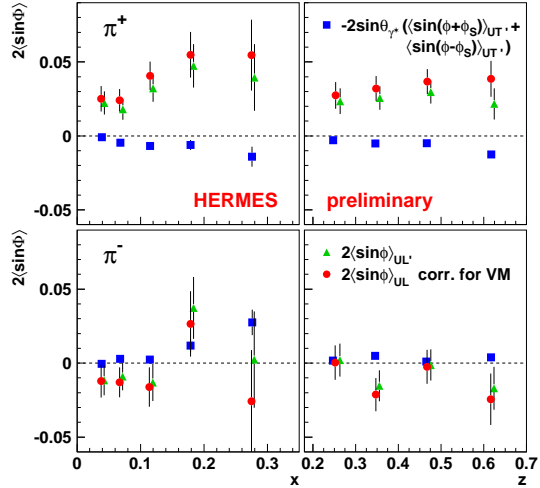}}
  \caption[]{The extracted subleading--twist component 
    $2\lmom\sin\phi\rmom_\mathrm{UL}$ ($\bullet$) and the measured
    moments $2\lmom\sin\phi\rmom_\mathrm{UL^\prime}$ and  $- 2 \sin\theta_{\gamma^*} 
    \left[\lmom \sin(\phi+\phi_S)\rmom_\mathrm{UT^\prime} +
      \lmom \sin(\phi-\phi_S)
      \rmom_\mathrm{UT^\prime}\right]$ for longitudinally ($\blacktriangle$) and 
    transversely ({\tiny$\blacksquare$}) polarised targets, respectively.\\[-2\baselineskip]}
  \label{AUL}
\end{figure}

With measurements on longitudinal and transverse targets it is now possible to extract
the subleading--twist contribution to the cross section:
\begin{equation}
  \lmom\sin\phi\rmom_\mathrm{UL} = \lmom\sin\phi\rmom_\mathrm{UL^\prime}
  + \sin\theta_{\gamma^*} \left[\lmom \sin(\phi+\phi_S)\rmom_\mathrm{UT^\prime}
  + \lmom \sin(\phi-\phi_S)\rmom_\mathrm{UT^\prime}\right]\ ,
  \label{eq:AUTl}
\end{equation}
which is valid up to corrections of order $\sin^2\theta_{\gamma^*}$ \cite{Die05}.
$\sin\theta_{\gamma^*}$ can be calculated from the lepton kinematics.
In Fig.~\ref{AUL} the extracted subleading--twist components are shown for different
kinematic bins together with the longitudinal asymmetry moments of reference \cite{Air00}, 
which were reanalysed to have the same binning in $x$ and $z$. In addition, the sum of 
the measured transverse asymmetry moments multiplied by the suppression factor 
$\sin\theta_{\gamma^*}$ (see Eq.~(\ref{eq:AUTl})) is plotted. The overall systematic 
uncertainty of the subleading--twist component $2\lmom\sin\phi\rmom_\mathrm{UL}$ is less 
than 0.003. For negative pions $2\lmom\sin\phi\rmom_\mathrm{UL}$ is compatible with zero,
for positive pions the subleading--twist component is positive and of the order of 2--5\%.
This is a strong indication for large subleading--twist DFs and FFs.
In case of $\pi^+$ they clearly dominate the results of the 
measurement with a longitudinally polarised hydrogen target. For both pion types the 
contribution from the transverse polarisation component is small.

One should note that also the measured transverse asymmetry moments have a small contribution
of the longitudinal polarisation component $S_\|=\sin\theta_{\gamma^*}\cos\phi_S$ which
couples to the $\sin\phi$ modulation of the longitudinal cross section. In order to estimate 
the subleading--twist contribution to the measured transverse asymmetry moments 
$2\lmom\sin(\phi\pm\phi_S)\rmom_\mathrm{UT^\prime}$,
the extracted subleading--twist component has to be multiplied by 
$\frac{1}{2}\sin\theta_{\gamma^*}$ \cite{Die05}. The resulting maximum 
contribution to the measured moments in reference \cite{Air05_2} amounts to an absolute value
of 0.004. For the measured transverse asymmetry moments in Fig.~\ref{AUT} that corresponds to a 
correction which is negligible compared to the statistical uncertainty.

\section*{Notes}
\begin{notes}
\item[a] E-mail: ulrike.elschenbroich@desy.de
\end{notes}

\vfill\eject
\end{document}